\documentclass[11pt]{article}
\usepackage[dvipdfm]{graphicx}
\usepackage{wrapfig, booktabs,amsmath,amsthm}
\usepackage{moreverb}
\newtheorem{theorem}{Theorem}
\newtheorem{lemma}[theorem]{Lemma}
\newtheorem{fact}[theorem]{Fact}
\begin{document}
\title{A Note on the Middle Levels Conjecture}
\author{Manabu Shimada$^*$ \qquad Kazuyuki Amano\footnote{Deptartment of
Computer Science, Gunma University, 1-5-1 Tenjin, Kiryu, Gunma 376-8515 Japan,
\tt{\{shimada@amano-lab|amano@\}.cs.gunma-u.ac.jp}}}
\maketitle
\section*{Abstract}
{\small
The middle levels conjecture asserts that there is a Hamiltonian 
cycle in the middle two levels of $2k+1$-dimensional hypercube.
The conjecture is known to be true for $k \leq 17$ 
[I.~Shields, B.J.~Shields and C.D.~Savage, Disc. Math., 309, 
5271--5277 (2009)].
In this note, we verify that the conjecture is also true for $k=18$ by
constructing a Hamiltonian cycle in the middle two levels of 
37-dimensional hypercube with the aid of the computer.
We achieve this by introducing a new decomposition technique and
an efficient algorithm for ordering the Narayana objects.
}


\section{Introduction}
Let $Q_n$ denote the $n$-dimensional hypercube, i.e., 
$Q_n$ is a graph with $2^n$ vertices, each vertex is labeled by 
an $n$-bit binary string and two vertices are adjacent iff their
strings differ exactly in one bit.
The $i$-th level of $Q_n$ is the set of vertices labeled by
strings with exactly $i$ ones.

\begin{figure}[h]
\begin{center}
\includegraphics[width=13cm ,clip]{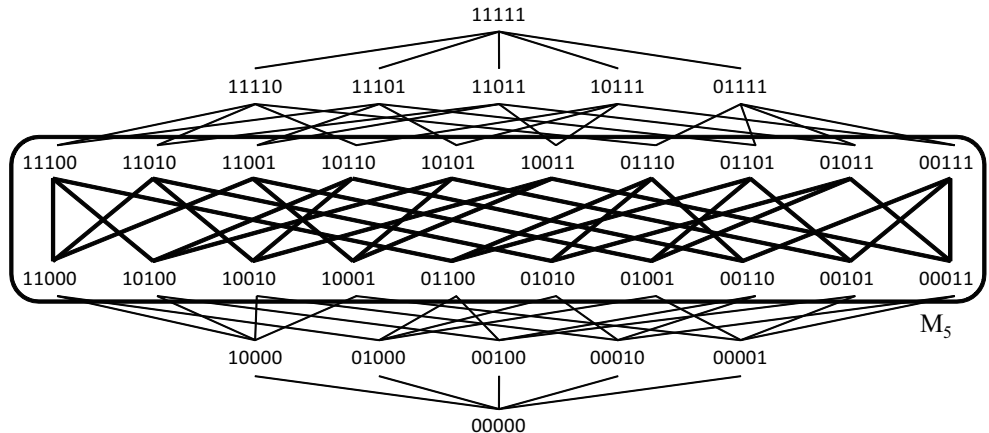}
\caption{The hypercube $Q_5$ and the middle levels graph $M_5$.}
\label{fg:Q5}
\end{center}
\end{figure}

The {\it middle levels graph} is a subgraph of $Q_{2k+1}$
induced by the middle two levels $k$ and $k+1$, and is denoted
by $M_{2k+1}$ (see Fig. \ref{fg:Q5}).
The {\it middle levels conjecture} asserts that
the graph $M_{2k+1}$ has a Hamiltonian cycle for every $k$.
It appears as an ``exercise" in Knuth's book 
\cite[Exercise 56, Sect. 7.2.1.3]{knuth}, in which
the conjecture is credited to Buck and Wiedermann \cite{BW84}.

In spite of considerable efforts, the conjecture remains open
(see e.g., \cite{john,shields2} and the references therein).
It was shown to be true for $k \leq 11$ by Moews and Reid, and
for $12 \leq k \leq 15$ by Shields and Savage \cite{shields1} and
$16 \leq k \leq 17$ by Shields et al. \cite{shields2}.

In this note, we verify that the conjecture is also true for $k=18$ by
constructing a Hamiltonian cycle in the middle two levels of 
37-dimensional hypercube with the aid of the computer.
We achieve this by plugging a new decomposition technique and
an efficient algorithm for ordering the Narayana objects
into a Hamiltonian path heuristic developed by Shields et al. 
\cite{shields2, shields1}.
In the largest case, our program could find a 
Hamiltonian path in
a graph with $\sim 2.36\cdot 10^8$ vertices in about a week on a
standard PC.

The organization of this note is as follows.
In Section \ref{reduce}, we briefly review the approach taken by 
Shields et al. \cite{shields2,shields1} for reducing the size of
the problem.
In Section \ref{run}, we describe an additional reduction
that decomposes the problem into a number of smaller subproblems.
In Section \ref{narayana}, we introduce an efficient algorithm for
ordering the Narayana objects which was helpful for reducing the 
resource needed in the computation. 
Finally in Section 5, we summarize our computational results.
Throughout the paper, $n=2k+1$ denotes the dimension
of a hypercube.


\section{Reducing the problem}\label{reduce}
The Hamiltonicity of the middle levels graph, which
has $2{n \choose k}$ vertices, can be reduced
to the problem for finding a suitable Hamiltonian path
in a smaller graph with ${n \choose k}/n$ vertices \cite{shields1}.

For an $n$-bit binary sequence $x=x_1x_2\cdots x_n$,
define the {\it cyclic shift} $\sigma$ by
$\sigma(x)=x_{2}x_{3}\cdots x_{n}x_{1}$.
For every two vertices $x$  and $y$ in $M_n$,
$x$ and $y$ are adjacent iff $\sigma(x)$ and $\sigma(y)$ are adjacent.
This naturally introduces an equivalence relation $\sim$ 
on the set of vertices of $M_n$ such that $x \sim y$ iff
$x = \sigma^i(x)$ for some integer $i$.
By noticing that that $\sigma^n(x)=x$ for every $x$, each equivalence
class has $n$ elements.

\begin{figure}
\begin{center}
\includegraphics[width=12cm ,clip]{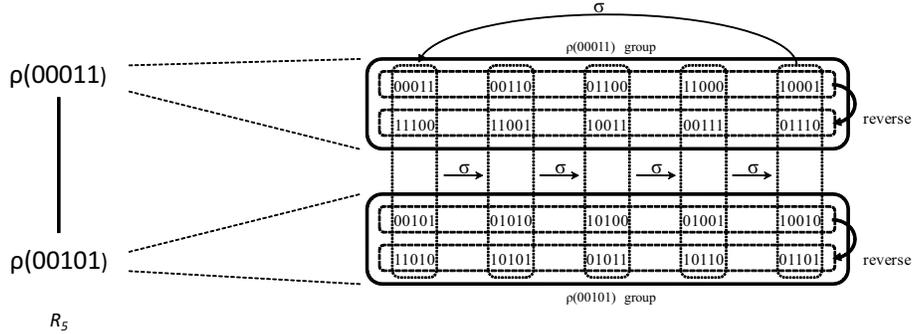}
\caption{The graph $R_5$ and its relationship to the vertices
of $M_5$.}
\label{fg:redu}
\end{center}
\end{figure}

A further reduction can be made by considering the {\it complement}.
The complement of an $n$-bit binary string $x=x_1x_2 \cdots x_n$ is
$\bar{x}=\bar{x_{1}}\bar{x_{2}}\cdots \bar{x_{n}}$.
Note that two vertices $x$ and $y$ are adjacent iff
$\bar{x}$ and $\bar{y}$ are adjacent.
By considering these two operations,
the vertices of $M_n$ is partitioned into $|M_n|/2n$ classes,
each of them has $2n$ vertices (Fig. \ref{fg:redu}).
Here and hereafter, we denote the number of vertices of a
graph $G$ by $|G|$.

For an $n$-bit binary sequence $x$,
let $\rho(x)$ denote this equivalence class including $x$, i.e.,
$\rho(x)=\{\sigma^i(x), \sigma^i(\bar{x}) \mid 
0 \leq i < n\}$. 
Let $R_n$ denote the graph whose vertices are these equivalence
classes and two vertices $\rho(x)$ and $\rho(y)$ in $R_n$ 
are adjacent iff there is an edge between $u$ and $v$ in $M_n$ for
some $u \in \rho(x)$ and $v \in \rho(y)$.

The following lemma, which was shown by Shields and Savage \cite{shields1}, guarantees that we can lift a Hamiltonian path in $R_n$ to
a Hamiltonian cycle in the middle levels graph.

\begin{lemma}
\label{Lem:shields}
If there is a Hamiltonian path in $R_n$ starting from the
vertex $\rho (0^{k+1}1^{k})$ and ending at the vertex 
$\rho (0(01)^{k})$, then there is a Hamiltonian cycle in
$M_n$.
\end{lemma}


\section{Decomposition based on Runs}\label{run}
Since the graph $R_n$ is still huge 
(i.e., $|R_{37}| \sim 4.8 \cdot 10^8$),
we divide $R_n$ into a number of smaller graphs and 
search them individually and possibly in parallel.

A {\it run} of a binary string $x$ is a consecutive appearance of
1's or 0's in $x$.
For example, we say that $000000$ has one run and $001011$ has four runs.
We will divide $R_n$ into three parts depending on the number
of runs of strings in a vertex.
Notice that $\rho(x)$ may contain strings having different runs.
We pick a string with $k$ one's such that it starts with $0$ and
ends with $1$ as a representative of $\rho(x)$, and the
number of runs of this string is referred as the
number of runs of $\rho(x)$.
Since this number is always even, we introduce a new unit
called ``brun" which is equal to two runs.

\begin{figure}
\begin{center}
\includegraphics[width=7cm ,clip]{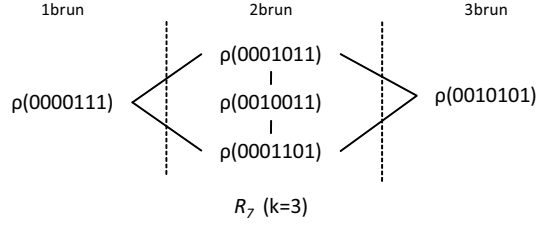}
\caption{The graph $R_7$ and the decomposition of $R_n$ 
based on ``brun".}
\label{fg:brun}
\end{center}
\end{figure}

\begin{figure}[b]
\begin{center}
\includegraphics[width=8cm ,clip]{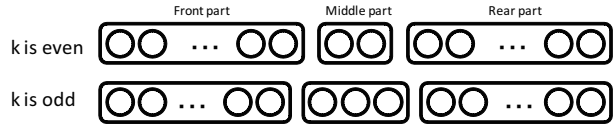}
\caption{The decomposition of $R_n$. Each small circle represents
an induced subgraph by the vertices with a specified brun.}
\label{fg:part}
\end{center}
\end{figure}

Note that, in $R_n$, only $\rho(0^{k+1}1^{k})$ has 1 brun and
only $\rho (0(01)^{k})$ has $k$ bruns.
In a preliminary experiment,
we found that a decomposition based on the following three intervals
is plausible (see Figs. \ref{fg:brun} and \ref{fg:part}).

\begin{itemize}
\item Front part : $1\sim (\lfloor k /2 \rfloor-1)$\ brun(s)
\item Middle part : $\lfloor k /2 \rfloor \sim 
(k - \lfloor k /2 \rfloor +1)$\ bruns
\item Rear part : $(k - \lfloor k/2 \rfloor +2)\sim k$\ bruns
\end{itemize} 
Note that, when $k=18$, these three intervals are 
$\{1,2, \ldots, 8\}, \{9,10\}$ and $\{11, \ldots, 18\}$.

We will find a Hamiltonian path in each of these three graphs and
then connect them to get a Hamiltonian path in $R_n$.
In order to apply Lemma \ref{Lem:shields},
we fix the start vertex of a path in the front part to 
$\rho(0^{k+1}1^{k})$ and the end vertex of a path in the rear part to 
$\rho (0(01)^{k})$.
In addition, we should satisfy the additional requirements that
(i) an end vertex of a path in the front part is adjacent to a 
start vertex of a path in the middle part, and 
(ii) an end vertex of a path in the middle part is adjacent to a
start vertex of a path in the rear part.

After some considerations, we pick strings
$hc(k,r) := 0^{k-r+1}(01)^{r}1^{k-r}$ as terminals of paths.
Note that $\rho (hc(k,r))$ has a maximum number of neighbors
in vertices with $r-1$ bruns and with $r+1$ bruns, respectively.
In addition,
(i) $hc(k,1) = 0^{k+1}1^{k}$,
(ii)$hc(k,k) = 0(01)^{k}$, and
(iii) for every $i$, $\rho (hc(k,i))$ and $\rho (Rev(hc(k,i+1)))$ are 
adjacent in $R_n$ where $Rev(x)$ denotes the reverse of a string 
$x=x_1x_2\cdots x_n$ i.e., $Rev(x)=x_n\cdots x_2 x_1$.
We also use the following fact which can easily be verified.

\begin{fact}
\label{Fact:rev}
Let $\{\ell,\ell+1, \ldots, r\}$ be a subset of 
$\{1,2, \ldots, k\}$.  Suppose that there is a Hamiltonian path
in an induced subgraph of $R_n$ with vertices of at least
$\ell$ bruns and at most $r$ bruns that starts from $\rho(x)$ and ends
at $\rho(y)$.
Then there is a Hamiltonian path in the same graph that
starts from $\rho (Rev(x))$ and ends at $\rho (Rev(y))$.
\end{fact}

For a Hamiltonian path $P$, let $Rev(P)$ denote a Hamiltonian path
in a same graph whose existence is guaranteed by Fact \ref{Fact:rev}.
In summary, 
our search procedure is the following:
First find a Hamiltonian path in each of three parts of the graph
starting from $\rho(hc(k,\ell))$ and ending at 
$\rho(hc(k,r))$ where $\ell$ and $r$
are the left-end and right-end of each interval, and
let denote these three paths as $P_{F}$, $P_{M}$ and $P_{R}$.
Then connect $P_{F}$, $Rev(P_{M})$ and $P_{R}$ in this order to
get a Hamiltonian path in $R_n$ which fulfills the condition in Lemma 
\ref{Lem:shields}.


\section{Ordering of Vertices}\label{narayana}
Each vertex of the graph $R_n$ can naturally be stored using 
$n$ bits of memory.
However, this can be reduced by using an efficient ordering
of the vertices.
Indeed, since the number of vertices of $R_n$ is less than $2^{32}$
for $n \leq 39$, we can store them using a 32-bit integer par item.
In this section, we give an efficient algorithm for ordering
the vertices of our reduced graphs.
A bit surprisingly, plugging this ordering scheme into a 
program gives a significant improvement of a running time 
of the program that will be shown in the next section.

\subsection{View Vertices of Middle Levels as Catalan Objects}
The $n$-th Catalan number is the number of expressions 
containing $n$ pairs of
parentheses which are correctly matched and is well-known to be 
$$C(n)=\frac{1}{n+1}
\left( 
\begin{array}{c}
2n\\
n\\
\end{array} 
\right).
$$
Notice that the number of vertices in $R_n$ is
equal to the $k$-th Catalan number $C(k)$.
This suggests that there is a bijection
between the set of vertices of $R_n$ and the set of correctly matched
$n$ pairs of parentheses.

In the following, we identify a sequence of parentheses
with a binary string under a mapping ``(" $\leftrightarrow$
``0" and ``)" $\leftrightarrow$ ``1".
In addition, by a technical reason, we add one ``0" to the top of 
the string.
For example, we consider that ``(()(()))" represents the
string ``$000100111$''.
An $2k+1$ bit binary string starting with $0$ is 
said to be {\it correctly matched} if it is corresponding to a correctly 
matched $n$ pairs of parentheses.

\begin{fact} 
For every vertex $\rho(x)$ in $R_n$, there is a unique correctly
matched string in $\rho(x)$.
\end{fact}

\begin{proof}
We should only consider a string with $k$ one's since
no string with $k+1$ one's is correctly matched.

Suppose that we represent a string by a path in the grid 
such that it goes upward when we read 0 and downward when we read 1.
For example, a path for the string
$0000111$ is drawn as Fig. \ref{fg:0000111}.
It is clear that a string $x$ is 
correctly matched iff the starting point of the path for 
$x$ is located at the lowest level in the path 
and it is only the point on this level.
\begin{figure}[h]
\begin{center}
\includegraphics[width=4cm ,clip]{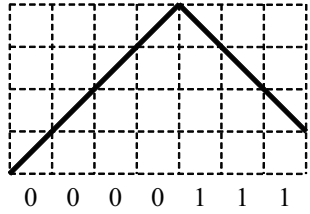}
\caption{A path for the string ``0000111".}
\label{fg:0000111}
\end{center}
\end{figure}

Recall that $\rho(x)$ contains every string that
obtained from $x$ by applying the cycle shift an arbitrary times.
Note that, for every $x$ with $k$ one's,
a path for $x$ ends at one step higher than
the starting point of the path.
Hence if we draw paths for $x$ and $\sigma^i(x)$
for some $i$, a path for the substring that 
shifted backward in $\sigma^i(x)$ is drawn at one level higher than 
the original level (Fig. \ref{fg:cycle}).

\begin{figure}[h]
\begin{center}
\includegraphics[width=7cm ,clip]{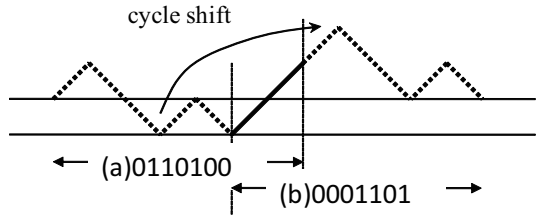}
\caption{A path for $x=0110100$ (a) and for 
$\sigma^5(x)=0001101$ (b). A dotted line represents a
path for the substring `01101' which goes backward by the 
cycle shift in (b).}
\label{fg:cycle}
\end{center}
\end{figure}

By this observation, it is easy to see that a 
correctly matched string in $\rho(x)$ can be obtained by 
(i) draw a path for $x$, and pick the rightmost point 
among all points on the lowest level of the path, and
(ii) shift $x$ so that this point becomes the top of the 
resulting string.

It is also easy to see that every other string in $\rho(x)$ is not
correctly matched. This guarantees the uniqueness and
hence completes the proof.
\end{proof}

By this fact, there is a bijection from 
the set of vertices in $R_n$ to the Catalan objects, i.e.,
the vertices in $R_n$ are uniquely mapped to integers
$\{0,1, \ldots, C(k)-1\}$.

\subsection{Lexicographical Ordering for Catalan Objects}
In our programs,
we number vertices $\rho(x)$ in $R_n$ according to the
lexicographical ordering (starting from 0) of a 
correctly matched string in $\rho(x)$.

Obviously, the ordering of a string $x$ is equal to the
number of strings lexicographically smaller than $x$.
Hence if we can count the number of strings smaller than $\tilde{x}$
for a given prefix $\tilde{x}$, then the ordering of $x$ can easily
be computed.
For example, the ordering of the string $0010101$ in a set 
$S \subseteq \{0,1\}^7$ can be computed as the sum of
the numbers of strings in $S$ starting from $000$, 
$00100$ and $0010100$.

\begin{figure}
\begin{center}
\includegraphics[width=5cm ,clip]{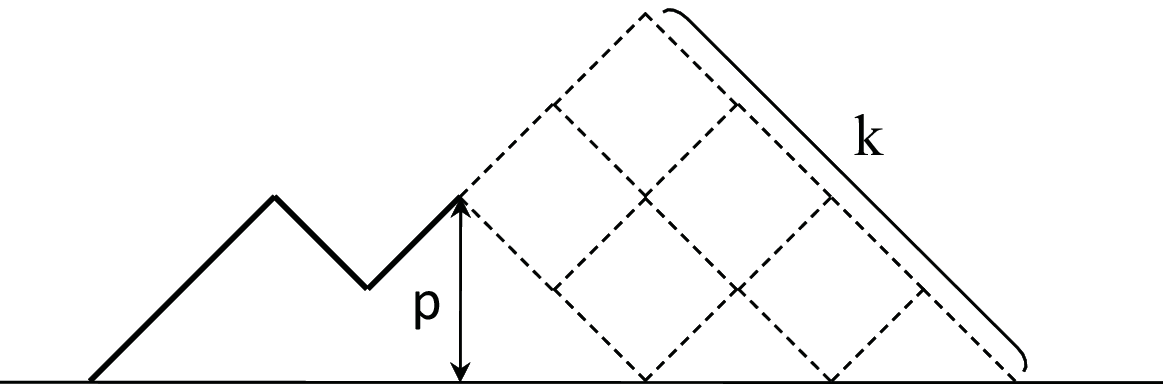}
\caption{$C_w(k,p)$ is equal to the 
number of left-right paths in the grid.}
\label{fg:weight}
\end{center}
\end{figure}

Let $P_\ell \subseteq \{0,1\}^{2\ell+1}$ be the set of correctly
matched strings of length $2\ell+1$.
For a prefix $\tilde{x} \in \{0,1\}^t$ with $t \leq 2\ell+1$,
the number of strings in $P_\ell$ starting with $\tilde{x}$ is shown to be
\begin{eqnarray}
\label{Eq:cat}
C_w(k,p) =\frac{p+1}{k+1}
\left( 
\begin{array}{c}
2k-p\\
k-p\\
\end{array} 
\right),
\end{eqnarray}
where $p=\sharp_0(\tilde{x})-\sharp_1(\tilde{x})-1$
and $k=\ell-\sharp_1(\tilde{x})$.
Here we denote the number of $0$'s and $1$'s
in $\tilde{x}$ by $\sharp_0(\tilde{x})$ and $\sharp_1(\tilde{x})$,
respectively.
Intuitively, $p$ denotes the height of the end point of a path 
for $\tilde{x}$ and $k$ denotes the number of ``remaining" one's 
in a string (see Fig. \ref{fg:weight}).
Note that these numbers are known as the Catalan Triangle
(see e.g., the sequence A009766 of \cite{web1}).
Using Eq. (\ref{Eq:cat}),
we can calculate the lexicographical ordering of a
vertex $\rho(x)$ efficiently.
For example, the ordering of $\rho(0010101)$ is given
by $C_w(3,2)+C_w(2,2)+C_w(1,2)=3+1+0=4$.

\subsection{Runs and Narayana Numbers}
Since we decompose the graph $R_n$ into smaller parts,
it is desirable to give an efficient ordering
algorithm for the set of vertices of these decomposed graphs.
By a similar argument to that in Section 4.1,
the number of vertices of $R_n$ with $r$ bruns
is shown to be
$$N(k,r)=\frac{1}{k}
\left(\begin{array}{c}k\\r\\ \end{array}\right)
\left(\begin{array}{c}k\\r-1\\ \end{array}\right),$$
which is known as the Narayana numbers.
$N(k,r)$ is the number of correctly matched
$k$ pairs of parentheses that contains the 
subsequence ``()" exactly $r$ times.
Note that the Catalan numbers are represented by the sum of
the Narayana numbers, i.e.,
$$C(k) = \sum^{k}_{i=1}N(k,i).$$

It is also shown that the lexicographical ordering of a string
$x$ in the set of correctly matched strings with $r$ bruns can
be efficiently computed using the following formula:
$$N_w(k,p,r)=\frac{k+(p-1)(r-1)}{k(k-p-r+1)}
\left(\begin{array}{c}k-p\\r\\ \end{array}\right)
\left(\begin{array}{c}k\\r-1\\ \end{array}\right),$$
that represents the number of correctly matched strings 
of which the meanings of $p$ and $k$ are the same as 
in Eq. (\ref{Eq:cat})
and $r$ denotes the `remaining' number of the subsequence ``()".
A detailed discussion on how to compute the ordering for 
such Narayana objects will be appeared in the full version of 
this note.


\section{Computational Results}
We develop a program for finding a Hamiltonian path
for decomposed graphs based on the algorithm proposed by
Shields et al.\cite{shields1}
in which we represent the vertices of graphs by the ordering
described in Section \ref{narayana}.
Using this program, we have succeeded to find a desired
Hamiltonian path for every three parts, i.e., the front, middle, rear 
parts of $R_n$ for every $8 \leq k \leq 18$, which shows the 
Hamiltonicity of the middle levels graphs for $k\leq 18$.
Note that, for smaller values of $k$, our decomposition schema would not
work.

The computational results are summarized in Table \ref{result}.
Our program is executed on a PC with
an Intel Xeon processor of $2.26$ GHz and $24$ GB of memory available.
Note that the maximum memory used in our experiments was about $9$ GB.
We show the elapsed time in seconds, and the case
that takes less than 1 second is shown as 0.

The second column
shows the elapsed time of a base program to find a path in the
entire graph $R_n$.
In a base program, we don't use our ordering scheme and vertices
are stored as $n$-bit strings.
The third column
shows the longest elapsed time of a base program
for finding a path in each of three decomposed graphs.
The fourth column
shows the elapsed time of a program with the ordering technique
for the entire graph $R_n$.
The later columns show the elapsed time of a program
in which both techniques, i.e., the decomposition described
in Section 3 and the ordering described in Section 4.3 are 
included.

\begin{table}[ht]
\begin{center}
\caption{}
Running time to find a Hamiltonian cycle in the middle levels graph
\label{result}
\begin{tabular}{cccccccc}\toprule
k&Base&w/Decomp.&w/Ordering&\multicolumn{4}{c} {w/Decomp.+Ordering} \\ 
&&&&Front&Middle&Rear&Max\\\midrule
8	&	0	&	0	&	0	&	0	&	0	&	0	&	0	\\
9	&	0	&	0	&	0	&	0	&	0	&	0	&	0	\\
10	&	0	&	0	&	0	&	0	&	0	&	0	&	0	\\
11	&	1	&	1	&	1	&	0	&	1	&	0	&	1	\\
12	&	7	&	5	&	6	&	2	&	4	&	1	&	4	\\
13	&	51	&	45	&	30	&	2	&	22	&	3	&	22	\\
14	&	542	&	290	&	182	&	40	&	71	&	26	&	71	\\
15	&	7,657	&	3,003	&	1,984	&	133	&	477	&	83	&	477	\\
16	&	88,795	&	29,948	&	17,130	&	3,143	&	2,762	&	1,785	&	3,143	\\
17	&	-	&	542,821	&	195,330	&	15,226	&	25,329	&	6,410	&	25,329	\\
&&(6.3 days)&(2.3 days)\\
18	&	-	&	-	&	-	&	627,204	&	511,342	&	359,015	&	627,204\\
&&&&&&&(7.3 days)
\\\bottomrule
\end{tabular}
\end{center}
\end{table}
A bit surprisingly, introducing the ordering into a search
program gives a significant improvement of the running time.
The combination of our two techniques reduces the running time by
a factor of about 30 when $k=16$.
For $k=18$, the number of vertices of the front, middle 
and rear parts of the graph is 
$120,624,130$, $236,390,440$ and $120,624,130$, respectively.
Notice that the running time is the longest for the front part of
the graph.
This suggests that finding a Hamiltonian path is harder for a
graph consisting of vertices with smaller number of runs than that
with larger number of runs.

The source codes of the programs we used as well as some
additional data are available on the web page \cite{SA09}.
Note that our program can handle up to $k=19$.
At the time of writing this note, the search for the front and rear
parts of the graph for $k=19$ has been finished successfully, and that
for the middle part, which has about $1.18$ billion nodes,
is in progress.

\medskip
\noindent
{\bf Note Added}

Several months after writing the above, 
our program for finding a Hamiltonian path
in the middle part of the graph for $k=19$ has successfully terminated.
This confirms that the middle two levels of $39$-dimensional hypercube is
also Hamiltonian.
The number of vertices of the front, middle and rear parts of the
graph is about $2.92 \times 10^8$, $1.18 \times 10^{9}$ and
$2.92 \times 10^8$, respectively.
The running time of the program (executed on the same machine as above) is
about 56 days, 81 days and 27 days, respectively.

\end{document}